\documentclass{article}
\usepackage[preprint]{neurips_2026}

\usepackage{amsmath,amsfonts,bm}

\def\eqref#1{equation~\ref{#1}}
\def\1{\bm{1}}

\DeclareMathAlphabet{\mathsfit}{\encodingdefault}{\sfdefault}{m}{sl}
\SetMathAlphabet{\mathsfit}{bold}{\encodingdefault}{\sfdefault}{bx}{n}

\usepackage{hyperref}
\usepackage{url}
\usepackage{mathrsfs}

\usepackage{booktabs}
\usepackage{array}
\usepackage{float}
\usepackage{amssymb}
\usepackage{mathtools}
\usepackage{verbatim}
\usepackage{graphicx}
\usepackage{multirow}

\newcommand{\dd}{\mathrm{d}}
\newcommand{\btheta}{\theta}
\newcommand{\bx}{\mathbf{x}}
\newcommand{\tpi}{\pi}
\newcommand{\bz}{\mathbf{z}}
\newcommand{\nitems}{N_{\textrm{items}}}
\newcommand{\nresponses}{N_{\textrm{levels}}}

\title{Bayesian model-averaging stochastic item selection for adaptive testing}

\author{%
  Tianyi Su \\
  UCLA Department of Statistics \\
  Los Angeles, CA 90095 \\
  \texttt{kristinas7796@g.ucla.edu} \\
  \And
  Edison M.\ Choe \\
  Renaissance Learning \\
  Wisconsin Rapids, WI 54494 \\
  \texttt{edison.choe@renaissance.com} \\
  \And
  Joshua C.\ Chang \\
  NIH Clinical Center \\
  \texttt{josh.chang@nih.gov} \\
}

\begin{document}

\maketitle

\begin{abstract}
  Computer Adaptive Testing (CAT) aims to accurately estimate an individual's ability using only a subset of an Item Response Theory (IRT) instrument.
  Many applications also require diverse item exposure across testing sessions, preventing any single item from being over- or underutilized.
  In CAT, items are selected sequentially based on a running estimate of a respondent's ability.
  Prior methods almost universally see item selection through an optimization lens, motivating greedy item selection procedures.
  While efficient, these deterministic methods tend to have poor item exposure.
  Existing stochastic methods for item selection are ad-hoc, with item sampling weights that lack theoretical justification.
  We formulate stochastic CAT as a Bayesian model averaging problem.
  We seek item sampling probabilities, treated in the long-run frequentist sense, that perform optimal model averaging for the ability estimate in a Bayesian sense.
  The derivation yields an information criterion for optimal stochastic mixing: the expected entropy of the next posterior.
  We tested our method on seven publicly available psychometric instruments spanning personality, social attitudes, narcissism, and work preferences, in addition to the eight scales of the Work Disability Functional Assessment Battery.
  Across all instruments, accuracy differences between selection methods at a given test length are varied but minimal relative to the natural noise in ability estimation; however, the stochastic selector achieves full item bank exposure, resolving the longstanding tradeoff between measurement efficiency and item security at negligible accuracy cost.
\end{abstract}

\section{Introduction}

The combination of Item Response Theory (IRT) and Computer Adaptive Testing (CAT) forms the dominant methodology backing the use of exams for ability assessment.
High-profile examples of this pairing include the Graduate Management Admission Test (GMAT)~\citep{kingstonExploratoryStudyApplicability1985,rudnerImplementingGraduateManagement2009}, the nursing National Council Licensure Examination (NCLEX)~\citep{wooEnsuringValidityNCLEX2012}, the National Registry of Emergency Medical Technicians (NREMT)~\citep{venturaTakingNREMTExam2021}, and the Armed Services Vocational Aptitude Battery (ASVAB)~\citep{segallDevelopmentComputerizedAdaptive1999}.
IRT/CAT also features in many healthcare contexts because of its adaptation in Patient Reported Outcomes Measurement Information System (PROMIS) instruments~\citep{cellaFutureOutcomesMeasurement2007,cellaPatientReportedOutcomesMeasurement2010,segawaComparisonComputerAdaptive2020} that are widely used in FDA-regulated trials.

\textbf{Item Response Theory (IRT).}
IRT is a generative latent-variable framework that models how a respondent of ability $\theta$ responds to each item in a testing bank~\citep{bockIRTEstimationDomain1997,immekusMultidimensionalItemResponse2019}.
A large pool of items is administered to a calibration sample, and fitting the IRT model jointly determines each item's parameters and each respondent's ability~\citep{kieftenbeldRecoveryGradedResponse2012,burknerBayesianItemResponse2020,lordMaximumLikelihoodEstimation1983}.
To score a new respondent, we hold item parameters fixed and solve an inverse problem for the ability parameter.
The item bank ideally covers the entire range of possible abilities, but administering a large bank is burdensome.
Near any fixed ability, however, only a small number of items are informative---a fact that CAT exploits.

\textbf{Computer Adaptive Testing (CAT).}
CAT estimates a respondent's ability by sequentially selecting the most relevant items from a possibly large item battery.
Each candidate item is judged by how much it would improve the fidelity of the ability estimate.
The most common approach selects items greedily---an efficient procedure that nevertheless suffers from poor item exposure.

Item exposure refers to the rate at which individual items in a testing bank are presented across multiple administrations.
When exposure is poor, the effective instrument administered by the CAT is a limited subset of the items in the original bank, and the remaining items are wasted.
Poor exposure has two consequences.
First, combined with commonly-used improper scoring rules, it biases the resulting ability estimates.
Second, and more critically for high-stakes testing, predictable item selection compromises test security: if the same small set of items is administered to every examinee, those items can be memorized and shared, allowing bad actors to achieve high scores without possessing the underlying ability.
Maintaining broad item exposure is therefore essential for the integrity of any repeatedly administered adaptive test.

Because the running ability estimate is unreliable at the beginning of a test, the statistical measures used to compare candidate items are noisy.
Simply choosing the item that appears statistically best (a ``greedy'' approach) may therefore not be ideal.
A more effective strategy may be to hedge, selecting items that are useful across a wider range of potential ability levels.

We provide a methodology for hedging based on viewing item selection as a model selection problem.
Each candidate item implies a different model for the respondent's ability at the next step of the test.
Viewing the problem through the lens of model averaging, we derive an item selection criterion that corresponds to optimal ensembling in the space of ability estimates.

\textbf{Datasets.}
\emph{Public psychometric instruments:}
To evaluate our methodology across a diverse range of instruments, we additionally tested on seven publicly available psychometric scales: the Grit Scale (GRIT), the Right-Wing Authoritarianism scale (RWA), the Empathy Quotient / Systemizing Quotient (EQSQ), the Narcissistic Personality Inventory (NPI), the Work Preference Inventory (WPI), and the Time Management Assessment (TMA).
These instruments vary in the number of items, response cardinalities.
For each instrument, we fitted a Bayesian graded response model (GRM)~\citep{samejimaEstimationLatentAbility1969} using variational inference, and separate pairwise models used for ensembling and imputation.

\emph{Work Disability Functional Assessment Battery (WD-FAB):}
Consists of eight independent IRT models~\citep{meterkoWorkDisabilityFunctional2015,marfeoDevelopmentNewInstrument2016,marfeoMeasuringWorkRelated2019,changRegularizedBayesianCalibration2022,marfeoImprovingAssessmentWork2018a,jetteWorkDisabilityFunctional2019,porcinoWorkDisabilityFunctional2018}.
The item banks consist of questions that ask about a range of everyday activities.
Accepted responses were graded on either four or five option ordinal Likert scales.

\section{Preliminaries}

\begin{table}[t]
  \caption{Notation for all mathematical symbols.}
  \label{tab:notation}
  \centering
  \begin{tabular}{@{}ll@{}}
    \toprule
    \textbf{Parameters}                          & \textbf{Description}                                                         \\
    \midrule
    $\nitems$; $\nresponses$                     & Number of items; ordinal response categories                                 \\
    $\boldsymbol{x}$                             & Responses for all items (none missing)                                       \\
    $\textbf{x}_t$                               & Responses observed at step $t$                                               \\
    $\theta$, $\hat{\theta}_t$                   & Ability, and its expectation at step $t$                                     \\
    $w_{ik}$                                     & Weights used to calculate expected Fisher's Information                      \\
    $\textbf{z}_t$                               & Responses that have not yet been observed at step $t$                        \\
    $p^{(t)}(i)$                                 & Probability of choosing item $i$ at step $t-1$                               \\
    \midrule
    \textbf{Functions/operators}                 & \textbf{Description}                                                         \\
    \midrule
    $\pi(\theta|\textbf{x})$                     & Posterior density of $\theta$ given observed responses $\bx$                 \\
    $\pi(\theta)$                                & Prior density for $\theta$                                                   \\
    $\pi(x_i|\theta)$                            & Likelihood function; probability                                             \\
    $I_i(\theta)$                                & Expected Fisher Information at $\theta$                                      \\
    $\mathbb{E}$, $\textrm{Var}$,  $\mathcal{H}$ & Expectation, Variance, entropy                                                \\
    $\mathcal{D}$, $\mathcal{D}_x$               & KL divergence for continuous (ability) and discrete (response) distributions \\
    \bottomrule
  \end{tabular}
\end{table}

Consider a test bank of $\nitems$ items with a calibrated IRT model.
The model implies that a person of ability $\theta$ responds to the $i$-th item according to the probability mass function $\pi(x_i=k | \theta)$.
For generality, we assume a polytomous model with $\nresponses$ possible responses per item.
If calibration used a fully Bayesian approach, $\pi$ is the marginal probability mass obtained by integrating out the posterior item parameters; otherwise, it is the probability mass implied by point estimates.
Given a fully observed individual with responses $\boldsymbol{x} = (x_1,x_2,\ldots, x_{\nitems})$, we estimate ability by computing the statistics of the posterior distribution
\begin{equation}
  \pi(\theta | \boldsymbol{x} ) \propto \pi(\theta)\prod_{i=1}^{\nitems} \pi(x_i | \theta)
  \label{eq:full_posterior}
\end{equation}
where the maximum likelihood estimate corresponds to the mode when using a uniform $\pi(\theta).$
The objective of a testing session is to efficiently approximate the statistics of Eq.~\ref{eq:full_posterior}, the \emph{true posterior estimate} of a respondent's ability.
In CAT, items are presented sequentially.
At step $t$, $t$ items have been answered, yielding a running ability estimate.
Letting $\bx_t$ and $\bz_t$ denote the observed and unobserved responses at step $t$, we obtain the ability estimate by marginalizing over the unobserved responses:
\begin{equation}
  \pi(\theta | \bx_t) = \sum_{\bz_t} \pi(\theta, \bz_t|\bx_t) = \sum_{\bz_t} \pi(\theta | \bx_t, \bz_t) \, \pi(\bz_t | \bx_t). \label{eq:marginal_ability}
\end{equation}
The distribution $\pi(\bz_t | \bx_t) = \int \pi(\bz_t | \theta) \pi(\theta | \bx_t) \dd\theta$ couples the unobserved responses to the ability estimate through the model.
When the IRT model is correctly specified, the unobserved items are conditionally independent given $\theta$ and therefore ignorable; the marginalization collapses to the product $\pi(\theta)\prod_{i\in\bx_t} \pi(x_i | \theta)$, which is the standard ability estimate used in virtually all CAT implementations.
However, when the model is misspecified, ignoring the unobserved responses introduces bias~\citep{rubinInferenceMissingData1976}.
We retain the general form (Eq.~\ref{eq:marginal_ability}) and introduce an imputation model $\pi^\star(\bz_t|\bx_t)$ to handle the marginalization, reducing to the standard estimate as a special case.
The next item is then selected conditional on the resulting ability estimate.

\textbf{Prior art.}
The oldest and perhaps most-popular CAT methodology selects the item with the maximum local \emph{Fisher information} at the current point estimate $\hat\theta_t$~\citep{magisNoteEquivalenceObserved2015}.
The Fisher information selector has several known limitations.
First, it adjudicates items conditional on $\hat\theta_t$, which is poorly characterized early in an exam.
A class of modifications take ability uncertainty into account by computing a \emph{Bayesian Fisher information}---the expectation of the Fisher information over the current ability posterior~\citep{owenBayesianSequentialProcedure1975,vanderlindenBayesianItemSelection1998,vanderlindenFastSimpleAlgorithm2020,uenoAdaptiveTestingBased2013,choiComparisonCATItem2009}.
Information theoretic alternatives include the \emph{global information} method of \citet{changGlobalInformationApproach1996}, which we show decomposes into an expected KL divergence between successive ability estimates plus a discrete divergence term (Appendix~\ref{sec:global_info}).
Other related KL-based criteria exist~\citep{sorrelAdaptingCognitiveDiagnosis2020,wangItemSelectionMultidimensional2011,weissmanMutualInformationItem2007,wangNoteRelationshipShannon2020}.

Second, the Fisher information provides an inaccurate approximation of precision when few items have been observed.
One may instead directly compute the item-specific conditional \emph{Bayesian variance} $\textrm{Var}[\theta | \bx_t, \alpha_{t+1} = i]$~\citep{vanderlindenBayesianItemSelection1998}.

Third, greedy item selection methods have highly stereotypical item trajectories and poor item exposure.
Explicit exposure controls exist~\citep{georgiadouReviewItemExposure2007,hanComponentsItemSelection2018}, including stochastic methods that sample items according to an ad-hoc function of the Fisher information~\citep{barradaIncorporatingRandomnessFisher2008}, but these require tuning a temperature parameter and adaptive dampening to balance randomization against efficiency.

These prior methods target reduction of the posterior variance but do not consider whether the resulting ability estimate is well-calibrated.
\citet{zhuangBoundedAbilityEstimation2023} introduced a gradient-based method that selects items matching the gradient of the likelihood at an estimate of the true full-bank ability.
In Appendix~\ref{sec:loo}, we show that our criterion admits a leave-one-out cross-validation interpretation: we select the item whose omission would yield the largest deviation from the full-bank posterior.

CAT can be viewed as a particular application of Bayesian Optimal Experimental Design (BOED), which is a broad framework for choosing the next \emph{experiment} or measurement for learning about a system based on maximizing a given utility~\citep{rainforthModernBayesianExperimental2023}.
Unsurprisingly, many of the methods common to CAT have analogues in BOED, for instance in using Bayesian information theoretic criteria~\citep{sebastianiMaximumEntropySampling2000,bernardoExpectedInformationExpected1979} or Frequentist experiment/itemwise Fisher information~\citep{smithStandardDeviationsAdjusted1918}.
The most common criterion in modern BOED is the expected information gain (EIG), and many stochastic methods for approximating this quantity exist~\citep{lainez-aguirreStochasticProgrammingApproach2015,fosterUnifiedStochasticGradient2020,zaballaStochasticGradientBayesian2023,godaUnbiasedMLMCStochastic2022}.
However, unlike in CAT, there is not a strong motivation to use stochastic selection in order to improve exposure for BOED experiments.

\section{Methods}
Our main theoretical contribution is to frame CAT through the lens of model averaging rather than optimization.
Relative to a true data generation process $\mathcal{G}$, we define the information loss of a model $\mathcal{M}$ using the discrepancy $\mathcal{D}(\mathcal{G},\mathcal{M}).$
Given multiple candidate models $\mathcal{M}_1,\ldots,\mathcal{M}_n,$ the optimal~\citep{bozdoganModelSelectionAkaikes1987} ensemble model (that minimizes information loss) averages predictions from the candidate models according to weights
\begin{align}
  w_i \propto \exp(-\mathcal{D}(\mathcal{G},\mathcal{M}_i)).
  \label{eq:optimalweights}
\end{align}
Variants of discrepancy-weighted ensembling are common in prediction tasks.
Since $\mathcal{G}$ is unknown, cross validation~\citep{yaoUsingStackingAverage2018} and information criteria such as the AIC~\citep{akaikeLikelihoodTimeSeries1978,dormannModelAveragingEcology2018,wagenmakersAICModelSelection2004} approximate the information loss for a given model.

%In standard prediction, models have already been fitted to a dataset.
%In CAT, each candidate item implies a prospective ability model, and we develop methodology to perform information-loss-based weighting on future unobserved responses to determine the sampling probabilities for the next item.
%
At a given step $t$, the choice of the next item is analogous to choosing among $\nitems-t$ choices for the next ability estimate $\tpi(\theta|\bx_{t+1})$.
If we knew the full-bank estimate, we could compute the item-specific ability model discrepancy
\begin{equation}
  \mathcal{D}\left(\pi(\theta|\boldsymbol{x}) \parallel \tpi(\theta|\bx_{t+1}) \right)  = \int\pi(\theta|\boldsymbol{x})\log\frac{\pi(\theta|\boldsymbol{x})}{\tpi(\theta|\bx_{t+1})}\dd\theta = \textrm{const} - \int\pi(\theta|\boldsymbol{x})\log{\tpi(\theta|\bx_{t+1})}\dd\theta  \label{eq:objective},
\end{equation}
where the constant full-bank entropy term does not depend on item ordering, and use it as the basis for item selection.

As in all CAT methods, the objective (Eq.~\ref{eq:objective}) must be resolved under incomplete observation.
Dropping the constant portion we take the expectation of the second term over the unobserved responses using an imputation model $\pi^\star(\bz_t|\bx_t)$.
When the IRT model is correctly specified, $\pi^\star$ can be taken as the model-implied marginal $\pi(\bz_t|\bx_t) = \int \pi(\bz_t|\theta)\pi(\theta|\bx_t)\dd\theta$.
However, in the $\mathcal{M}$-open setting where the scoring model may be misspecified, $\pi^\star$ can be a more flexible imputation ensemble~\citep{yaoUsingStackingAverage2018}.
Taking the expectation:
\begin{align}
  &-\mathbb{E}_{\pi^\star(\bz_t|\bx_t)}\int\pi(\theta|\boldsymbol{x})\log \pi(\theta|\bx_{t+1})\dd\theta =  -\sum_{\bz_{t}} \pi^\star(\bz_t|\bx_t) \int \pi(\btheta|\bx_t,\bz_t)\log \pi(\theta|\bx_{t+1})\dd\theta \nonumber \\
  \MoveEqLeft\qquad\qquad\qquad\qquad\qquad\qquad\qquad = -\sum_{x_{t+1}}\!\int\pi^\star(x_{t+1}|\bx_t)\,\pi(\theta|x_{t+1},\bx_t)\log\pi(\theta|\bx_{t},x_{t+1})\dd\theta. \label{eq:expect}
\end{align}
Since items are selected one at a time, we evaluate the criterion per item -- the standard greedy process for pre-existing CAT methods.
Each item produces an ability model with the associated expected information loss (modulo additive constant)
\begin{align}
  \Delta^{(i)}_t = \sum_{k=1}^{\nresponses} \pi^\star(x_i=k|\bx_t)\,\mathcal{H}\!\left(\pi(\theta|\bx_t, x_i=k)\right),\label{eq:criterion}
\end{align}
where $\mathcal{H}(q) = -\int q(\theta)\log q(\theta)\,\dd\theta$ is the entropy and $\pi^\star(x_i=k|\bx_t)$ is the imputation model's predictive probability for item $i$.
The criterion of Eq.~\ref{eq:criterion} is the expected entropy of the next posterior: items that produce more concentrated posteriors have lower $\Delta^{(i)}_t$ and receive higher selection weight.
For a Gaussian posterior, $\mathcal{H}(\pi) = \frac{1}{2}\log\det(2\pi e\,\Sigma)$, so minimizing $\Delta^{(i)}_t$ is asymptotically equivalent to D-optimal design and subsumes both Fisher information and Bayesian variance as special cases.
The finite-sample advantage is that Eq.~\ref{eq:criterion} operates on the exact posterior and uses $\pi^\star$ rather than the model's own predictions to weight hypothetical responses.
Using Eq.~\ref{eq:criterion} in Eq.~\ref{eq:optimalweights}, the next item is sampled with probability $p^{(t+1)}(i) \propto \exp(-\Delta_t^{(i)})$ among the unadministered items $i \in \bz_t$.

\textbf{Numerical implementation.}
We coded two independent implementations of our methodology for the Graded Response Model: one in Python (using NumPy for vectorized tensor operations) and one in Golang (using only an ndarray library).
Neither implementation relies on specialized optimization or probabilistic programming libraries.
We approximated all integrals using the trapezoid rule with $200$ equally spaced grid points.

\section{Results}

\begin{table}[t]
  \caption{Summary of instruments used in evaluation.}
  \label{tab:datasets}
  \centering
  \small
  \begin{tabular}{@{}llrl@{}}
    \toprule
    \textbf{Instrument}     & \textbf{Abbreviation} & \textbf{Items} & $K$ \\
    \midrule
    \multicolumn{4}{@{}l}{\textit{OpenPsychometrics}} \\
    Sexual Compulsivity Scale           & SCS  & 10  & 4 \\
    Grit Scale                          & GRIT & 12  & 5 \\
    Generic Conspiracist Beliefs Scale  & GCBS & 15  & 5 \\
    Right-Wing Authoritarianism         & RWA  & 22  & 9 \\
    Narcissistic Personality Inventory  & NPI  & 40  & 2 \\
    Time Management Assessment          & TMA  & 50  & 2 \\
    Work Preference Inventory           & WPI  & 116 & 2 \\
    Empathy Quotient / Systemizing Quotient & EQSQ & 120 & 4 \\
    \midrule
    \multicolumn{4}{@{}l}{\textit{Work Disability Functional Assessment Battery (WD-FAB)}} \\
    \multicolumn{4}{@{}l}{8 unidimensional GRM scales (4 physical, 4 mental); $K \in \{4,5\}$} \\
    \bottomrule
  \end{tabular}
\end{table}

Table~\ref{tab:datasets} summarizes the instruments used in our evaluation.
In producing the following results, for each scale, we simulated item responses for $100$ respondents for each true underlying ability of $\theta\in\{-3, -2.5, -2,\ldots, 2.5, 3\}$, using a $\mathcal{N}(0,2)$ prior for ability scoring.
We then ran each respondent's item responses through each CAT item selection method, obtaining ability estimates at given test lengths.
The methods we evaluated are greedy selection via the Fisher information, Bayesian Fisher information, global information~\citep{changGlobalInformationApproach1996}, Bayesian variance, greedy discrepancy (Eq.~\ref{eq:criterion}), and our stochastic selection method ($p(i) \propto \exp(-\Delta_t^{(i)})$).
We also computed ability estimates for each simulated respondent based on all of their item responses, providing the ground-truth reference.

\begin{figure}[t]
\centering
\includegraphics[width=0.96\textwidth]{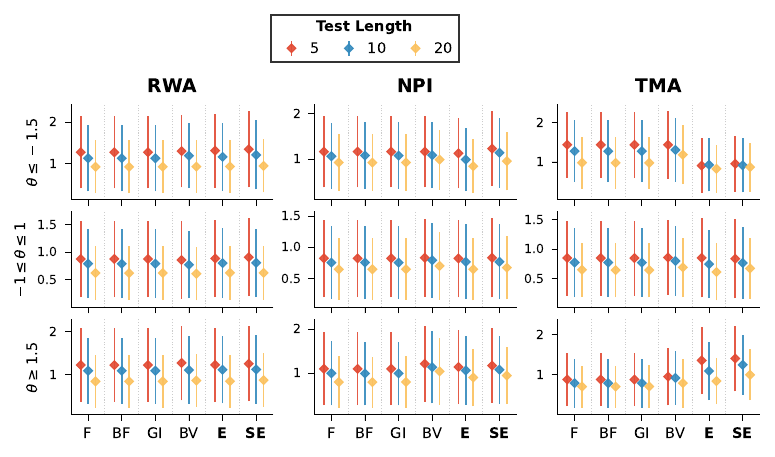}
\caption{\textbf{Absolute score error} $|\hat\theta_t - \hat\theta|$ (mean $\pm$ 1 SD) by ability range, dataset, and test length under $\mathcal{M}$-open simulation on OpenPsychometrics instruments.
Labels as in Fig.~\ref{fig:mopen_kl}. Lower is better.}
\label{fig:mopen_l2}
\end{figure}

\textbf{Openpsychometrics datasets.} 
For each dataset, we fitted a standard GRM and pairwise imputation models $\pi^\star$ in which every item serves as a univariate predictor of every other item.
We found optimal stacking weights by solving the constrained linear programming problem of~\citet{yaoBayesianHierarchicalStacking2022}.
These ensemble models better-approximate the original dataset than is capable of the GRM -- we used them to sample responses.
Results are broken down by ability range: high ($\theta \geq 1.5$), mid ($-1 \leq \theta \leq 1$), and low ($\theta \leq -1.5$).
All ability parameters are standardized so that $\theta = 1$ corresponds to one standard deviation in the calibration population.
Supplemental Figure~\ref{fig:mopen_kl} displays the KL divergence and Figure~\ref{fig:mopen_l2} displays the absolute score error, for each method, ability range, and test length.

All methods use the same baseline IRT scoring (without imputation) and the same $\mathcal{N}(0,2)$ ability prior, so the comparison isolates the item selection criterion.
The imputation model $\pi^\star$ is used only for generating $\mathcal{M}$-open responses---not for scoring---ensuring a fair comparison across selection methods.

\paragraph{Accuracy.}
Supplemental Tables~\ref{tab:mopen_kl} and~\ref{tab:mopen_l2} and Supplemental Figures~\ref{fig:mopen_kl} and~\ref{fig:mopen_l2} report the full accuracy results.
No method has a clear accuracy advantage across all datasets and test lengths.
Fisher, Bayesian Fisher, global information, and greedy entropy are nearly indistinguishable on most instruments. Where they differ, the difference is well within the level of natural noise.
The stochastic selector trades a small accuracy cost at short test lengths for near-identical performance at longer ones.

\begin{figure}[t]
\centering
\includegraphics[width=0.96\textwidth]{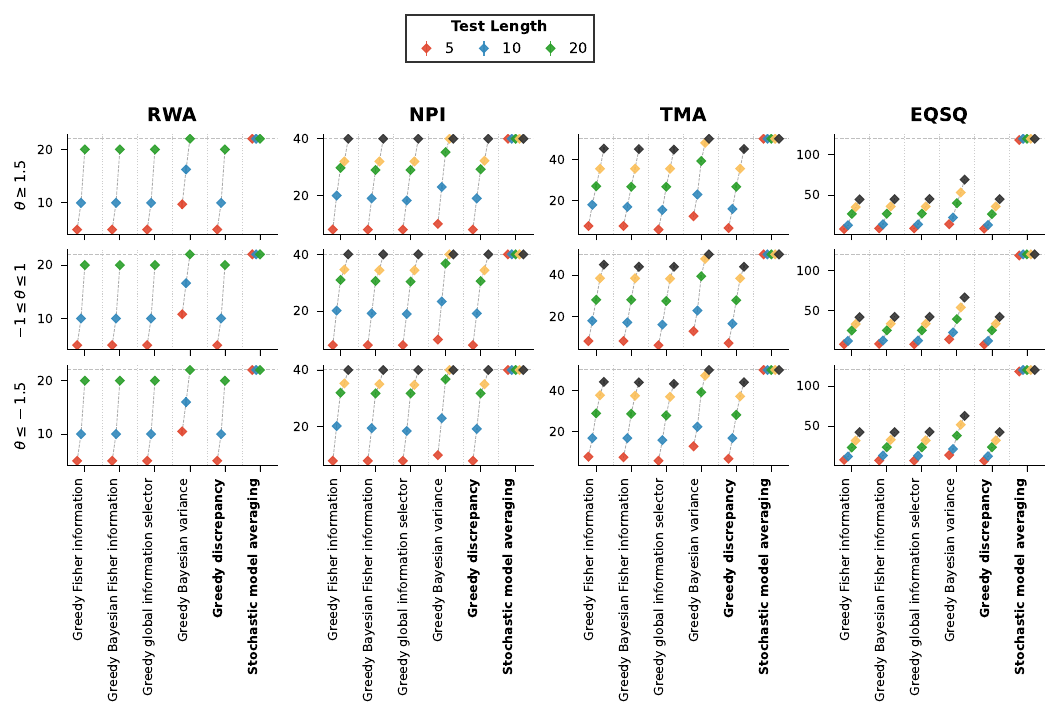}
\caption{\textbf{Item exposure} (unique items in union of 100 simulated CAT sessions) on OpenPsychometrics instruments, by ability range, method, and test length.
Dashed lines indicate total item bank size.
Methods labeled on the $x$-axis; \textbf{Greedy discrepancy} and \textbf{Stochastic model averaging} are the methods introduced in this work.
Higher is better.}
\label{fig:mopen_exposure}
\end{figure}

\paragraph{Exposure.} In contrast to the modest accuracy differences, the exposure differences are dramatic (Figure~\ref{fig:mopen_exposure}).
We computed item exposure as the number of unique items appearing in the union of item trajectories across 100 simulated CAT sessions at each ability level.
The greedy Fisher selector at $t{=}5$ uses exactly 5 items on RWA, 8 on NPI and TMA, and 8 on EQSQ---approximately $t$ items regardless of bank size.
The greedy entropy selector is similarly concentrated.
Even at $t{=}20$, Fisher exposes only 20 of 22 items on RWA, about 30 of 40 on NPI, 28 of 50 on TMA, and 25 of 120 on EQSQ.
The stochastic selector achieves complete or near-complete item bank coverage at every test length and ability range tested.
On RWA, NPI, and TMA, all items are exposed at $t{=}5$ across all ability ranges (22/22, 40/40, 50/50).
On the 120-item EQSQ, Fisher uses only 8 items at $t{=}5$ and 25 at $t{=}20$.

This exposure advantage comes with only a modest accuracy cost.
On TMA at $t{=}10$, the stochastic selector achieves full bank exposure while the accuracy difference is minimal (KL $2.22$ vs $2.19$ for Fisher; L2 $0.95$ vs $0.95$).
The practical implication is that test administrators can randomize item presentation---reducing memorization risk and increasing content coverage---without meaningfully degrading measurement precision.

\begin{figure}[t]
  \includegraphics[width=0.96\textwidth]{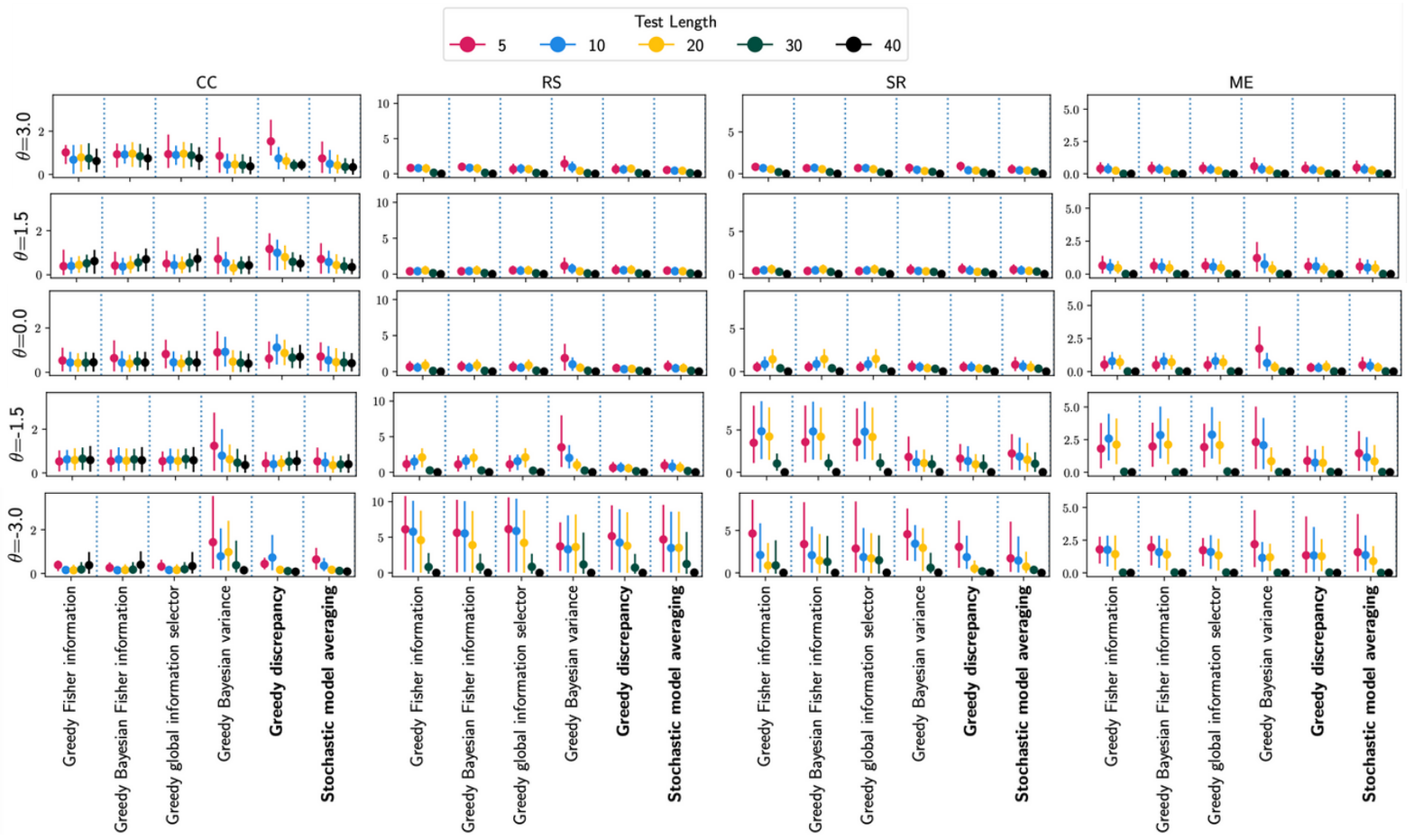}
  \caption{\textbf{Absolute error in ability estimate ($|\hat\theta_t - \hat\theta|$)} (mean and middle 80\% interval) conditional on true score $\theta$ by scale, item selection method, and test length $t$, for mental function scales of the WD-FAB. Lower is better.}
  \label{fig:absoluteerror}
\end{figure}

\textbf{WD-FAB results.} We used the same type of ensemble generating model for the WD-FAB as in the Openpsychometrics examples.
The KL divergence between CAT and full-bank ability estimates (Supplemental Fig.~\ref{fig:klerror}) reveals that Fisher information and global information selectors can \emph{increase} the discrepancy at intermediate test lengths, whereas Bayesian variance and our criterion (Eq.~\ref{eq:criterion}) reliably decrease it.
A selector that fails to decrease the discrepancy yields ill-calibrated ability estimates, compromising whole-distribution comparisons between individuals.

\begin{figure}[t]
  \includegraphics[width=0.96\textwidth]{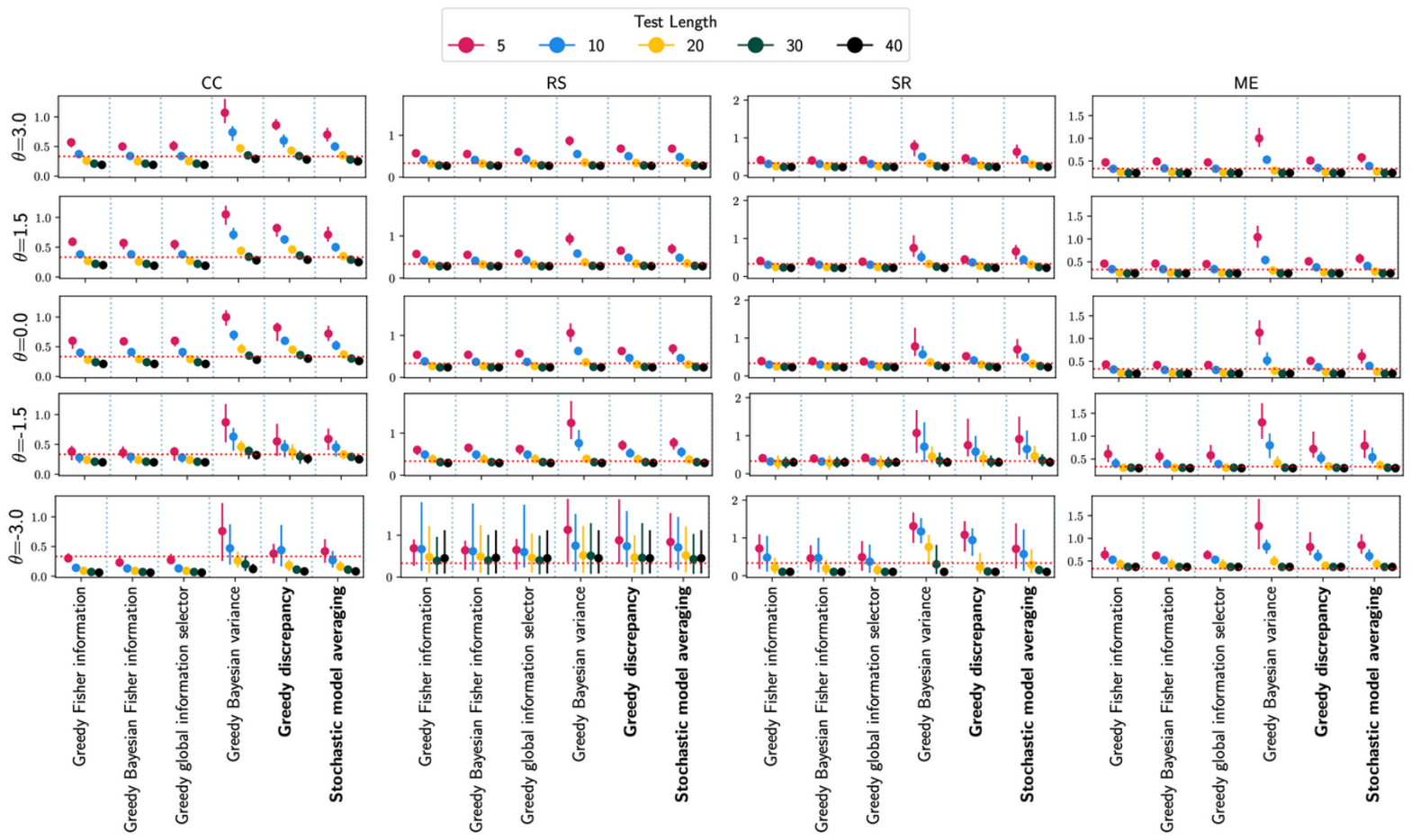}
  \caption{\textbf{Standard deviation of ability estimates ($\sqrt{\textrm{Var}_{t}(\theta)}$)} (mean and middle 80\% percentile) conditional on true score $\theta$ by scale and item selection method, for mental function scales of the WD-FAB. Used as stopping criteria for CAT. Lower is better.}
  \label{fig:ability_sd}
\end{figure}

Fig.~\ref{fig:absoluteerror} shows the absolute error of the posterior mean across simulations.
The error distributions are highly variable across scales.
Generally, error magnitude decreased as test length increased.
For most scales, a region of abilities exists where all selectors produce small errors.
No single method had the lowest errors in all situations, though the stochastic selector performed most consistently well.

The posterior variance often defines a stopping rule for CAT.
Fig.~\ref{fig:ability_sd} shows the standard deviation of posterior ability estimates across simulation configurations.
The two Fisher methods and the global information method yield the lowest posterior standard deviations.
However, Figures~\ref{fig:klerror} and~\ref{fig:absoluteerror} reveal that these estimates are ill-calibrated: the Fisher-based methods terminate prematurely and settle on sub-optimal ability estimates.

\begin{figure}[H]
  \includegraphics[width=\textwidth]{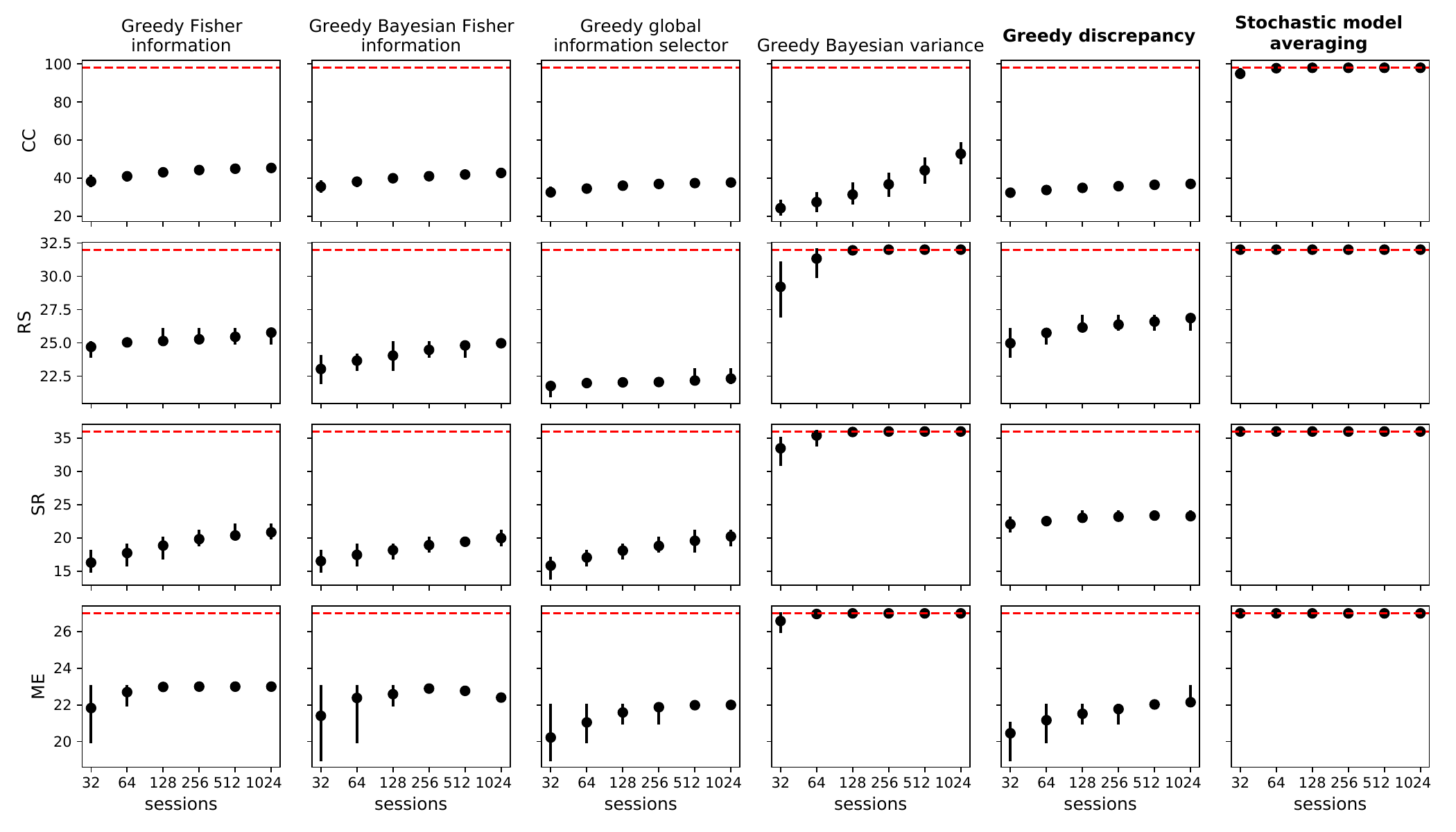}
  \caption{\textbf{Item exposure statistics} (mean and middle 80\% interval), for each of the given item selection methods across a given number of CAT sessions, for mental function scales of the WD-FAB. The dashed line represents the maximum possible exposure per scale. Higher is better.}
  \label{fig:testcoverage}
\end{figure}

\textbf{Item exposure (WD-FAB).} Fig.~\ref{fig:testcoverage} compares item exposure across sessions (12 items per scale, randomly distributed abilities).
We counted the number of unique items seen for each scale across replications of a given number of CAT sessions.
For example, on the ``ME'' scale, we estimate that 32 sessions expose approximately 22 items on average, though with wide variance.
As the number of sessions increases, so does the number of exposed items.
Among the greedy methods, Bayesian variance has the best item exposure, and on some scales it performs almost as well as our stochastic selector.
The stochastic selector successfully exposed all items for all scales in all scenarios we tested.

\section{Discussion}

We have introduced stochastic item selection for CAT derived from Bayesian model averaging.
Unlike prior stochastic methods~\citep{barradaIncorporatingRandomnessFisher2008} that require a temperature parameter and adaptive dampening, our sampling weights follow directly from discrepancy-weighted ensembling and have no hyperparameters to tune.

Across 15 instruments, we find that item selection method matters far less for accuracy than it does for exposure.
The stochastic selector achieves near-complete item bank coverage while incurring only a small accuracy cost, resolving the longstanding tradeoff between measurement efficiency and item security.
Bayesian variance can \emph{degrade} as the test lengthens under misspecification, particularly on TMA (Fig.~\ref{fig:mopen_kl}), whereas the entropy criterion remains stable.
On the WD-FAB, the Fisher information methods are over-confident in estimating scoring error (Fig.~\ref{fig:ability_sd}), settling on sub-optimal ability estimates.
The computational cost of our criterion is comparable to other Bayesian methods, as the shared hypothetical posterior updates dominate the per-item cost.

For  efficiency, randomization in CAT might seem sub-optimal.
Our simulations show that the efficiency cost is small.
Each criterion requires resolving unknown future responses, and since the true ability is unknown, the statistics of these responses are also unknown. 
So the greedy methods may be over-focused on optimizing a noisy objective.

\textbf{Limitations and extensions.}
We derive the marginal item probability mass functions in Eq.~\ref{eq:criterion} from the fitted IRT model.
More flexible IRT models~\citep{changProbabilisticallyautoencodedHorseshoedisentangledMultidomain2019,changAutoencodedSparseBayesian2023} could improve the accuracy of these expectations, provided that unobserved items are properly accounted for.
Although we formulate our methodology for a multidimensional ability parameter $\theta$, adapting the method to non-factorized multidimensional instruments would require additional work.
Balancing administration across different scales may require additional controls.

\subsubsection*{Acknowledgments}

This research was supported, in part, by the Intramural Research Program of the National Institutes of Health and the US Social Security Administration.
The views, information or content, and conclusions presented do not necessarily represent the official position or policy of, nor should any official endorsement be inferred on the part of, the Clinical Center, the National Institutes of Health, or the Department of Health and Human Services.

\bibliographystyle{plainnat}
\bibliography{irtvae}

\newpage
\appendix

\renewcommand{\thefigure}{S\arabic{figure}}
\renewcommand{\theequation}{S\arabic{equation}}
\setcounter{figure}{0}
\setcounter{equation}{0}

\section{Supplemental Methods}
\label{sec:supp_methods}

In this supplement we develop the mathematical relationships between our cross-entropy selection criterion (Eq.~\ref{eq:criterion}) and other selection criteria in the literature.
We show that the global information criterion decomposes into a term involving our discrepancy plus a discrete KL divergence, and that minimizing our discrepancy is equivalent to selecting the item whose omission would yield the largest leave-one-out discrepancy.

\subsection{Relationship to global information}
\label{sec:global_info}

The global information method of \citet{changGlobalInformationApproach1996} selects items by computing
\begin{align}
  \MoveEqLeft \textrm{Global information} = \mathbb{E}_\theta\left[ \sum_{k=1}^{\nresponses} \pi(x_i=k |\theta)\log\frac{\pi(x_i=k|\theta)}{\pi(x_i=k|\hat\theta_t)} \right] \nonumber                           \\
   & =\sum_{k=1}^{\nresponses} \int \tpi(\theta|\bx_t)\pi(x_i=k|\theta)\log\left[\frac{\pi(x_i=k|\theta)}{\pi(x_i=k |\hat\theta_t)}\frac{\tpi(\theta|\bx_t)}{\tpi(\theta|\bx_t)} \right]\dd\theta \nonumber      \\
   & = \sum_{k=1}^{\nresponses} {p}_{i}^{(t)}(k) \left[ \mathcal{D}\left( \tpi(\theta|\bx_t, x_{i}=k)\parallel \tpi(\theta|\bx_t)\right) - \log\pi(x_i=k |\hat\theta_t) +\log {p}_{i}^{(t)}(k) \right] \nonumber \\
   & =\mathbb{E}_{x_i}\left[\mathcal{D}\left( \tpi(\theta|\bx_t, x_i)\parallel \tpi(\theta|\bx_t)\right)  \right] +\mathcal{D}_{x_i}[{\pi}(x_i|\bx_t) \parallel \pi(x_i|\hat\theta_t)], \label{eq:gi_full}
\end{align}
where $\mathcal{D}(q(\theta)\parallel p(\theta))=\mathbb{E}_{q(\theta)} \log [q(\theta)/p(\theta)]$, $\mathcal{D}_{x}(p(x)\parallel q(x))=\sum_{k=1}^{\nresponses} p(k)\log(p(k)/q(k)),$
and $x_i\sim \pi(x_i|\bx_t)$ for
\begin{equation}
  {p}_{i}^{(t)}(k) = \pi(x_i=k|\bx_t) = \int \pi(x_i=k|\theta)\pi(\theta|\bx_t)\dd\theta.
\end{equation}

The first term in the last line of Eq.~\ref{eq:gi_full} is the expected KL divergence between the next-step ability estimate and the current ability estimate, taken over possible responses.
The second term is the discrete KL divergence between the marginal response distribution $\pi(x_i|\bx_t)$ (which integrates over the ability posterior) and the plug-in response distribution $\pi(x_i|\hat\theta_t)$ (which conditions on the point estimate).

Our cross-entropy criterion (Eq.~\ref{eq:criterion}) involves the cross-entropy $\mathcal{H}(\pi(\theta|\bx_t), \pi(\theta|\bx_t, x_i=k))$ rather than the KL divergence $\mathcal{D}(\pi(\theta|\bx_t, x_i=k) \parallel \pi(\theta|\bx_t))$.
The global information criterion thus conflates two distinct quantities: the informativeness of the item (first term) and the calibration of the point estimate (second term).
Our criterion, by contrast, directly measures the expected information loss relative to the full-bank posterior.

\subsection{Relationship to leave-one-out cross validation}
\label{sec:loo}

We can rewrite the discrepancy (Eq.~\ref{eq:objective}) to remove the explicit dependence on $\tpi(\theta|\bx_{t+1}),$
\begin{align}
  \MoveEqLeft \mathcal{D}\left(\pi(\theta|\boldsymbol{x})\parallel \tpi(\theta|\bx_{t+1}) \right) = \int\pi(\theta|\bx)\log\frac{\tpi(x_{t+1})\pi(\theta|\boldsymbol{x})}{\pi(x_{t+1}|\theta)\tpi(\theta|\bx_{t})}\dd\theta \nonumber \\
   & =\int \pi(\theta|\boldsymbol{x}) \log \frac{{p}_i^{(t)}(x_{t+1})}{\pi(x_{t+1}|\theta)}\dd\theta + \mathcal{D}(\pi(\theta|\boldsymbol{x})\parallel\tpi(\theta|\bx_t)) \label{eq:reparam}
\end{align}
where
$$
  {p}_{i}^{(t)}(k) = \int \pi(x_i=k|\theta)\tpi(\theta|\bx_t)\dd\theta,
$$
and note that while the second term in the last line of Eq.~\ref{eq:reparam} depends on the response for the next item, it does not depend on the choice of the next item.
We can then relate the discrepancy to leave-one-out (LOO) cross validation by expanding the first term in Eq.~\ref{eq:reparam}:
\begin{align}
  \MoveEqLeft \mathcal{D}\left(\pi(\theta|\boldsymbol{x})\parallel \tpi(\theta|\bx_{t+1}) \right) = \mathcal{D}(\pi(\theta|\boldsymbol{x})\parallel \tpi(\theta|\bx_t)) + \int \pi(\theta|\boldsymbol{x}) \log \frac{{p}_{i}^{(t)}(x_{i})\pi(\theta|\boldsymbol{x})}{\pi(\theta|\boldsymbol{x})\pi(x_{i}|\theta)}\dd\theta  \nonumber \\
   & =\mathcal{D}(\pi(\theta|\boldsymbol{x})\parallel \tpi(\theta|\bx_t))+S[\pi(\theta|\boldsymbol{x})] -\mathcal{D}(\pi(\theta|\boldsymbol{x})\parallel \tpi(\theta | \boldsymbol{x}\setminus\{x_i\})) + \log\frac{{p}_{i}^{(t)}(x_{i})}{{p}_i^{\textrm{LOO}}(x_{i})}
  \label{eq:reparam2}
\end{align}
where $\tpi(\theta | \boldsymbol{x}\setminus\{x_i\})$ is the ability estimate when leaving out $x_i$, which follows Bayes rule:
$$
  {\pi(x_i|\theta)\tpi(\theta | \boldsymbol{x}\setminus\{x_i\})} =  \pi(\theta|\boldsymbol{x}){{p}_{i}^{\textrm{LOO}}(x_i)},
$$
$S[\cdot]$ denotes the entropy, and ${p}_{i}^{\textrm{LOO}}(x_i) = \int \pi(x_i|\theta)\tpi(\theta | \boldsymbol{x}\setminus\{x_i\})\dd\theta$ is the corresponding LOO marginal mass function for item $i$.

In this representation, only the last two terms in Eq.~\ref{eq:reparam2} depend on the item choice.
The term $-\mathcal{D}(\pi(\theta|\boldsymbol{x})\parallel \tpi(\theta | \boldsymbol{x}\setminus\{x_i\}))$ measures the discrepancy between the full-bank estimate and the leave-one-out estimate---items whose omission causes large discrepancy are those that carry the most information about the ability.
The log-ratio term $\log({p}_{i}^{(t)}(x_{i})/{p}_i^{\textrm{LOO}}(x_{i}))$ compares the predicted response probability under the current running estimate to the LOO predictive probability, providing a calibration correction.

Therefore, minimizing the discrepancy is equivalent to selecting the item that, if left out of the full bank, would yield the largest discrepancy from the full-bank ability estimate.
This provides a cross-validation interpretation of our selection criterion: we are choosing the item that is most critical to the accuracy of the full-bank posterior.

\section{Supplemental Results: WD-FAB Physical Scales}

\begin{figure}[h!]
  \includegraphics[width=0.95\textwidth]{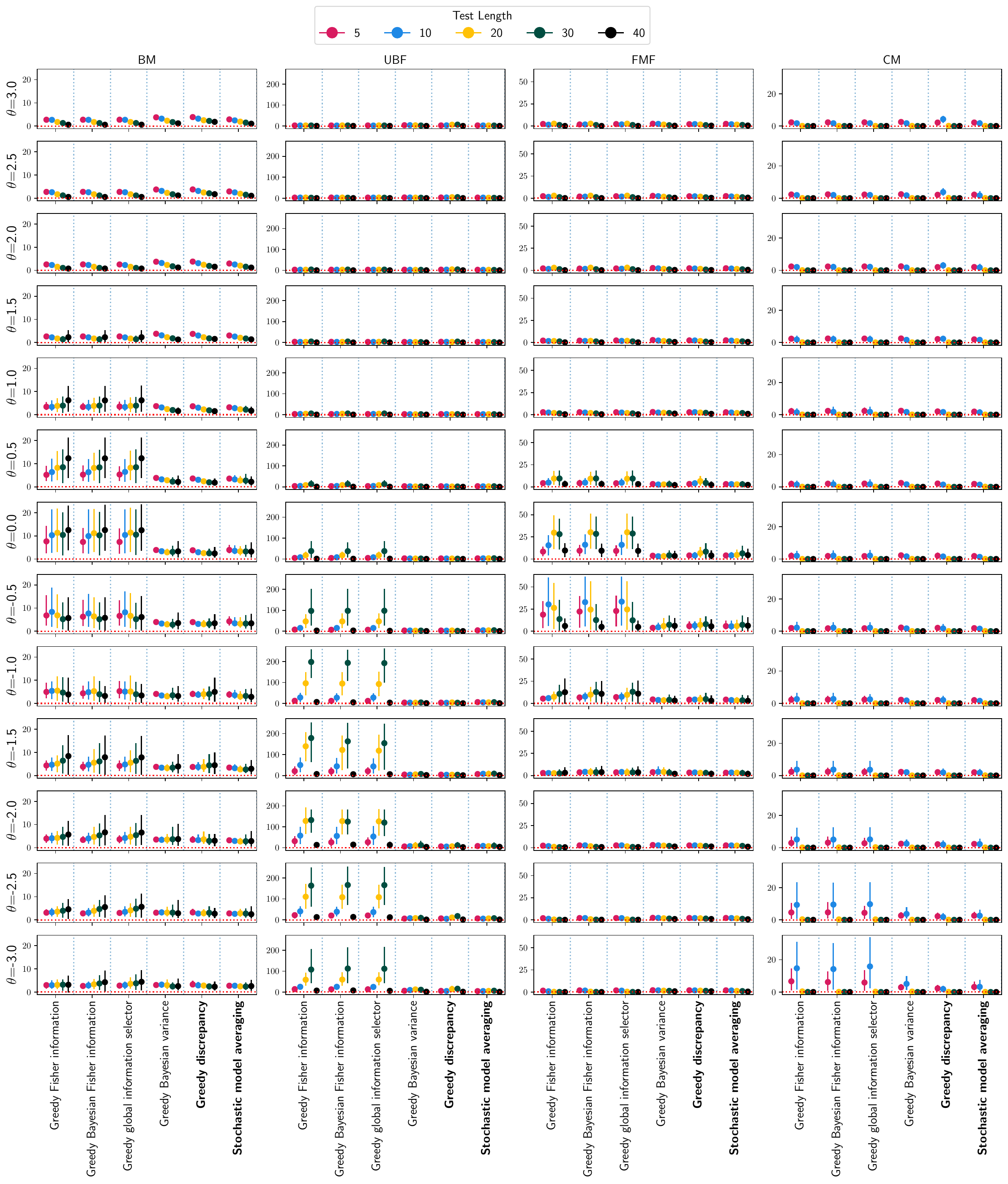}
  \caption{\textbf{Model discrepancy $\mathcal{D}(\pi(\theta|\bx) \parallel \tpi(\theta|\bx_t)$)} (mean and middle 80\% interval) conditional on score $\theta$ used to generate response sets, by scale, item selection method, and test length $t$, for physical function scales of the WD-FAB. Lower is better.}
  \label{fig:klerror_phys}
\end{figure}

\begin{figure}[h!]
  \includegraphics[width=\textwidth]{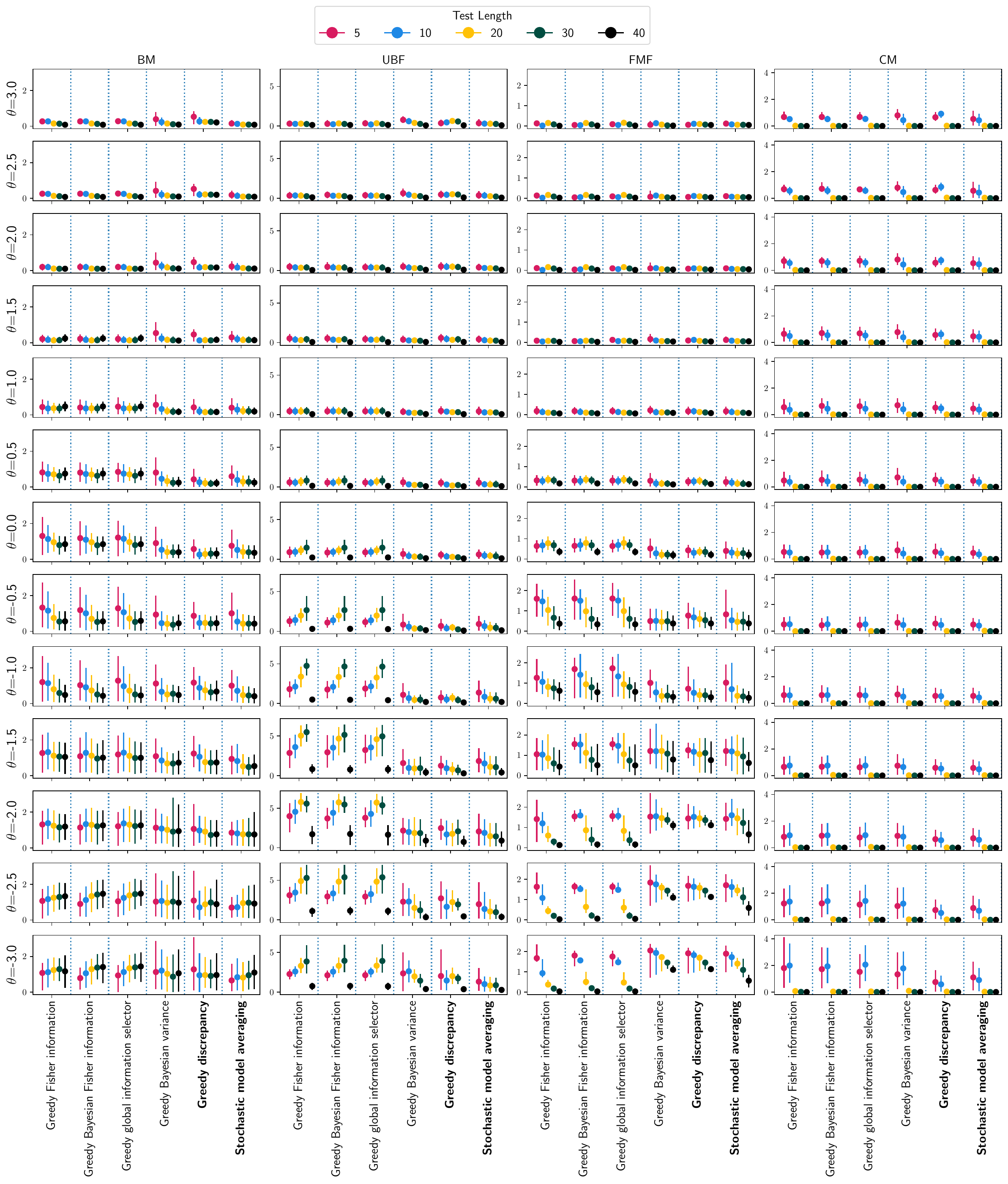}
  \caption{\textbf{Absolute error in means ($|\int\theta\tpi(\theta|\bx_t)\dd\theta - \int\theta\pi(\theta|\bx)\dd\theta|$)} (mean and middle 80\% interval) conditional on true score $\theta$ by scale, item selection method, and test length $t$, for physical function scales of the WD-FAB. Lower is better.}
  \label{fig:absoluteerror_phys}
\end{figure}

\begin{figure}[h!]
  \includegraphics[width=\textwidth]{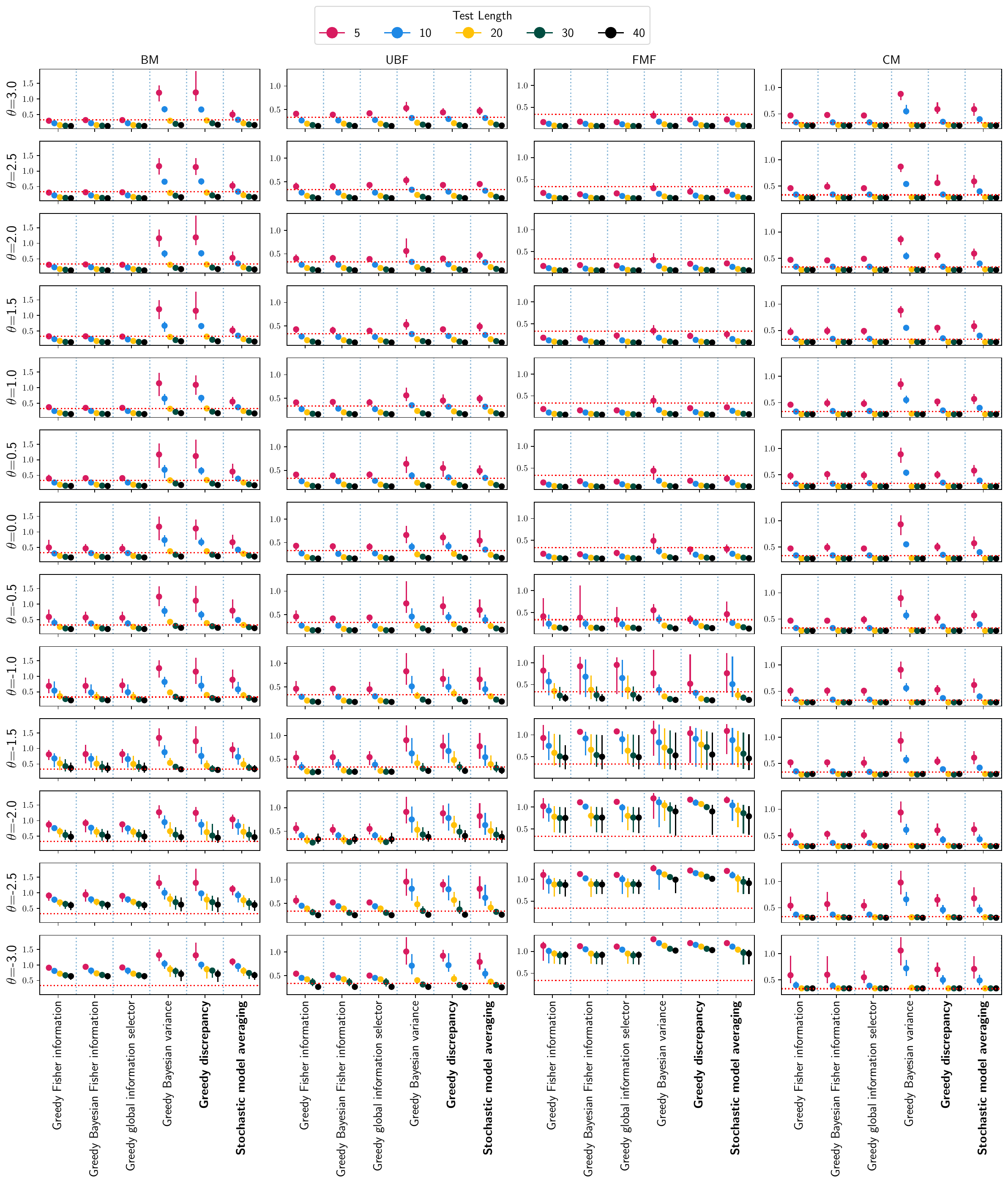}
  \caption{\textbf{Standard deviation of ability estimates ($\sqrt{\textrm{Var}_{t}(\theta)}$)} (mean and middle 80\% percentile) conditional on true score $\theta$ by scale and item selection method, for physical function scales of the WD-FAB. Used as stopping criteria for CAT.}
  \label{fig:ability_sd_phys}
\end{figure}

\begin{figure}[h!]
  \includegraphics[width=\textwidth]{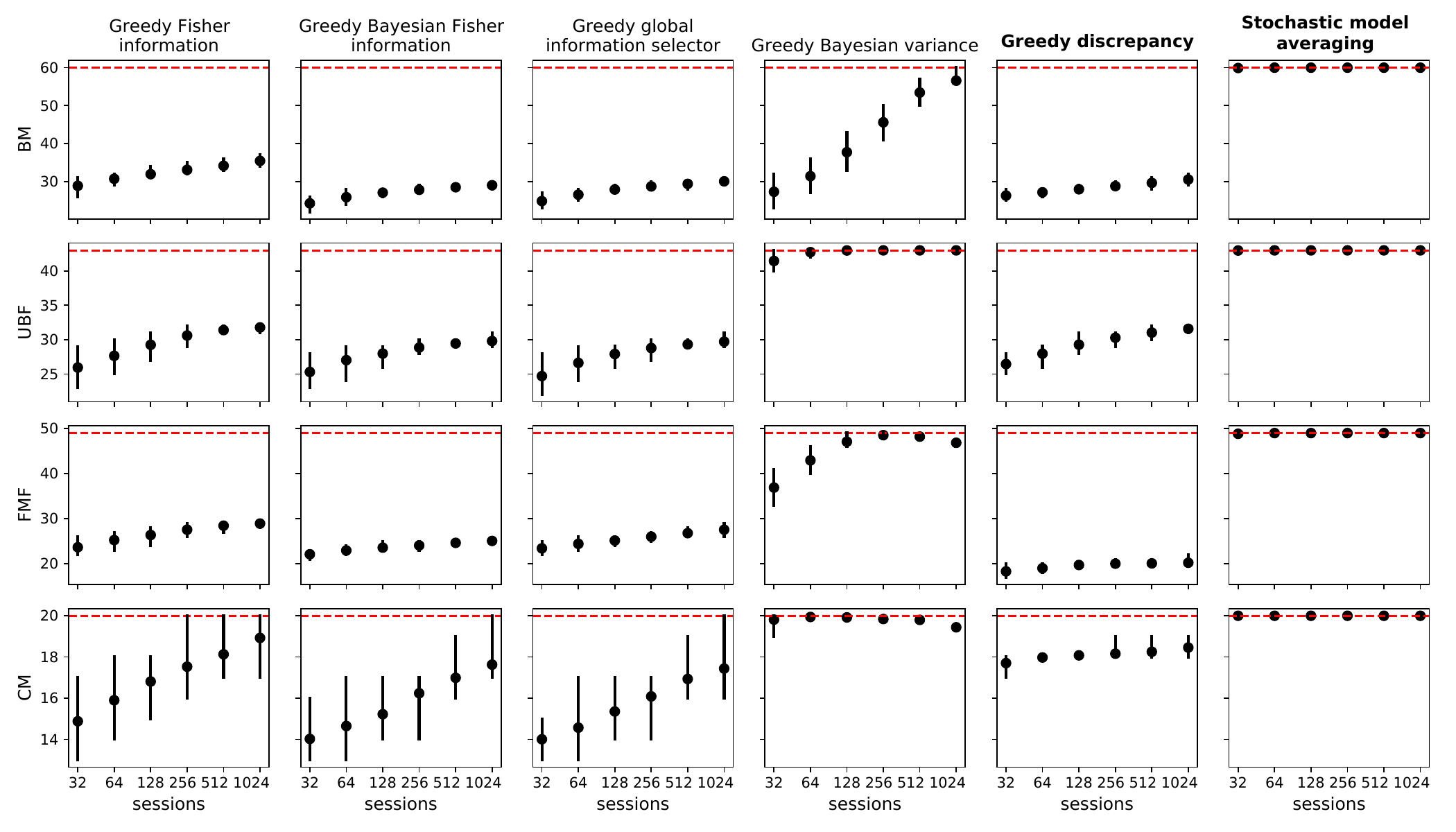}
  \caption{\textbf{Item exposure statistics} (mean and middle 80\% interval), for each of the given item selection methods across a given number of CAT sessions, for physical function scales of the WD-FAB. The dashed line represents the maximum possible exposure per scale. Higher is better.}
  \label{fig:testcoverage_phys}
\end{figure}

\section{Supplemental Results: KL Divergence Figures}

\begin{figure}[h!]
\centering
\includegraphics[width=0.96\textwidth]{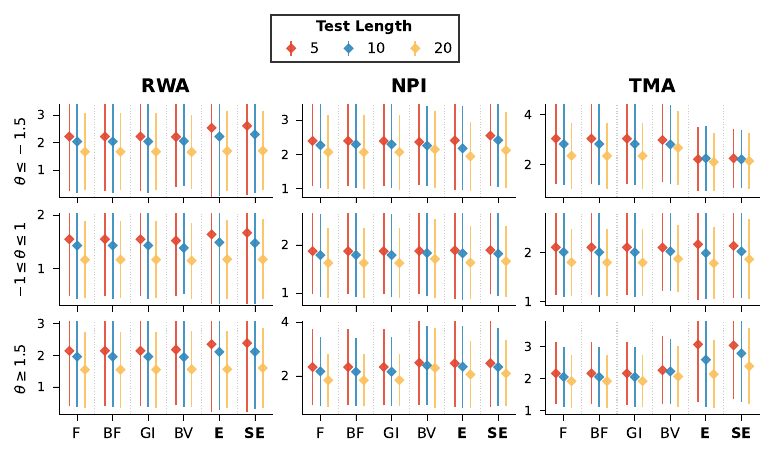}
\caption{\textbf{KL divergence} $\mathcal{D}(\pi(\theta|\bx) \| \tpi(\theta|\bx_t))$ (mean $\pm$ 1 SD) by ability range, dataset, and test length on OpenPsychometrics instruments.
Methods labeled on the $x$-axis; \textbf{Greedy discrepancy} and \textbf{Stochastic model averaging} are the methods introduced in this work.
All methods use the same baseline scoring.
Lower is better.}
\label{fig:mopen_kl}
\end{figure}

\begin{figure}[h!]
  \includegraphics[width=0.96\textwidth]{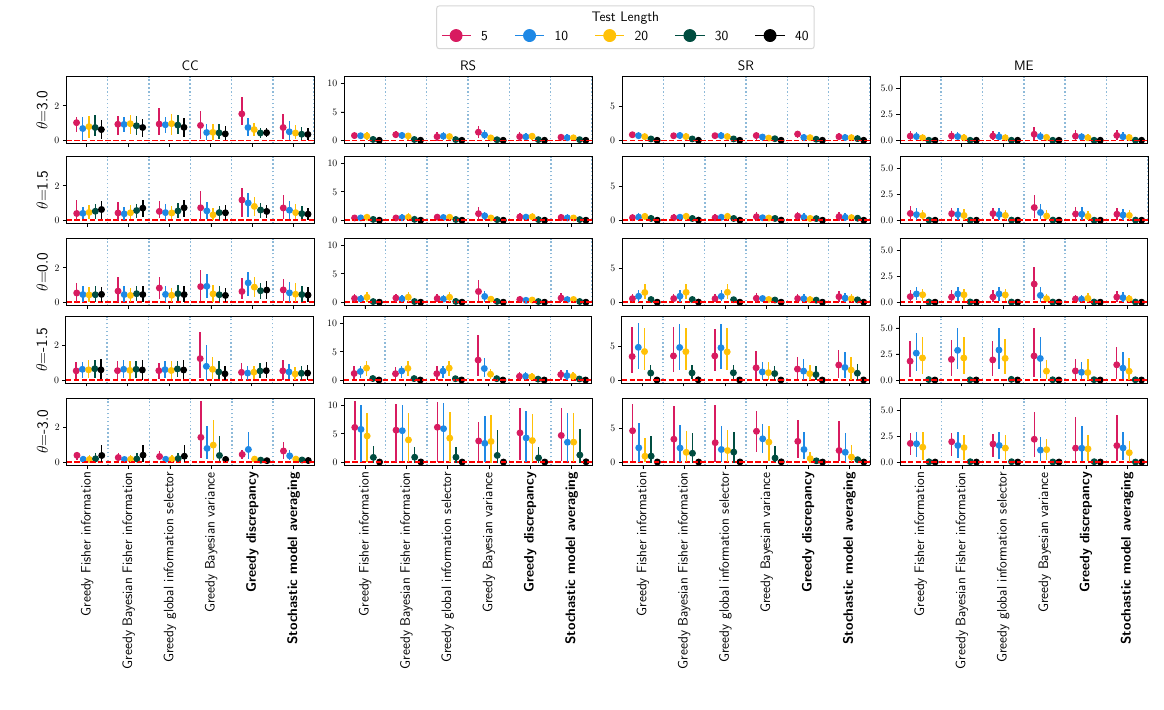}
  \caption{\textbf{Ability estimate discrepancy $\mathcal{D}(\pi(\theta|\bx) \parallel \tpi(\theta|\bx_t)$)} (mean and middle 80\% interval) conditional on score $\theta$ used to generate response sets, by scale, item selection method, and test length $t$, for mental function scales of the WD-FAB. Lower is better.}
  \label{fig:klerror}
\end{figure}

\section{Supplemental Results: OpenPsychometrics Tables}

\begin{table}[t]
\centering
\caption{\textbf{KL divergence} $\mathcal{D}(\pi(\theta|\bx) \| \tpi(\theta|\bx_t))$, mean (SD), under $\mathcal{M}$-open simulation.
Lower is better; \textbf{bold} = best per row. $\uparrow$ = increases with $t$ (pathological).
F = Fisher; BF = Bayesian Fisher; GI = global information; BV = Bayesian variance; E = entropy (greedy); SE = entropy (stochastic).}
\label{tab:mopen_kl}
\small
\begin{tabular}{@{}ll r r r r r r@{}}
\toprule
Dataset & $t$ & F & BF & GI & BV & E & SE \\
\midrule
  RWA & 5 & 2.33 (0.40) & 2.33 (0.40) & 2.33 (0.40) & 2.28 (0.40) & 2.33 (0.40) & \textbf{2.19 (0.28)} \\
  RWA & 10 & \textbf{1.93 (0.22)} & \textbf{1.93 (0.22)} & \textbf{1.93 (0.22)} & 1.99 (0.34) & \textbf{1.93 (0.22)} & 1.95 (0.23) \\
  RWA & 20 & \textbf{1.36 (0.13)} & \textbf{1.36 (0.13)} & \textbf{1.36 (0.13)} & \textbf{1.36 (0.15)} & \textbf{1.36 (0.13)} & 1.40 (0.16) \\
\midrule
  NPI & 5 & \textbf{2.13 (0.12)} & \textbf{2.13 (0.12)} & \textbf{2.13 (0.12)} & 2.14 (0.08) & \textbf{2.13 (0.12)} & 2.14 (0.09) \\
  NPI & 10 & \textbf{1.99 (0.11)} & 2.00 (0.12) & 2.00 (0.13) & 2.03 (0.06) & 2.00 (0.12) & 2.02 (0.09) \\
  NPI & 20 & \textbf{1.73 (0.07)} & \textbf{1.73 (0.07)} & \textbf{1.73 (0.07)} & 1.81 (0.08) & \textbf{1.73 (0.07)} & 1.76 (0.05) \\
\midrule
  TMA & 5 & 2.36 (0.10) & 2.35 (0.10) & 2.35 (0.10) & \textbf{2.34 (0.09)} & 2.35 (0.10) & \textbf{2.34 (0.09)} \\
  TMA & 10 & \textbf{2.19 (0.07)} & \textbf{2.19 (0.07)} & \textbf{2.19 (0.07)} & 2.24 (0.07) & \textbf{2.19 (0.07)} & 2.22 (0.07) \\
  TMA & 20 & \textbf{1.91 (0.03)} & 1.92 (0.03) & 1.92 (0.03) & 2.02 (0.06) & 1.92 (0.03) & 2.00 (0.05) \\
\midrule
  EQSQ & 5 & \textbf{3.25 (0.06)} & 3.28 (0.08) & 3.28 (0.08) & 3.32 (0.13) & 3.28 (0.08) & 3.28 (0.09) \\
  EQSQ & 10 & \textbf{3.12 (0.06)} & 3.13 (0.07) & 3.13 (0.07) & 3.20 (0.11) & 3.13 (0.07) & 3.17 (0.07) \\
  EQSQ & 20 & \textbf{2.92 (0.06)} & \textbf{2.92 (0.05)} & \textbf{2.92 (0.05)} & 3.00 (0.09) & \textbf{2.92 (0.05)} & 2.97 (0.06) \\
\midrule
  GRIT & 5 & \textbf{1.28 (0.17)} & \textbf{1.28 (0.17)} & \textbf{1.28 (0.17)} & 1.44 (0.36) & \textbf{1.28 (0.17)} & 1.37 (0.25) \\
  GRIT & 10 & \textbf{1.01 (0.17)} & \textbf{1.01 (0.17)} & \textbf{1.01 (0.17)} & 1.04 (0.21) & \textbf{1.01 (0.17)} & 1.02 (0.19) \\
\midrule
  SCS & 5 & 1.20 (0.14) & 1.20 (0.14) & 1.20 (0.14) & \textbf{1.13 (0.12)} & 1.20 (0.14) & 1.17 (0.14) \\
  SCS & 10 & \textbf{0.80 (0.08)} & \textbf{0.80 (0.08)} & \textbf{0.80 (0.08)} & \textbf{0.80 (0.08)} & \textbf{0.80 (0.08)} & \textbf{0.80 (0.08)} \\
\midrule
  GCBS & 5 & \textbf{1.58 (0.15)} & \textbf{1.58 (0.15)} & \textbf{1.58 (0.15)} & 1.64 (0.18) & \textbf{1.58 (0.15)} & 1.62 (0.16) \\
  GCBS & 10 & 1.28 (0.10) & \textbf{1.27 (0.10)} & \textbf{1.27 (0.10)} & 1.34 (0.15) & \textbf{1.27 (0.10)} & 1.30 (0.13) \\
\midrule
  WPI & 5 & 3.19 (0.06) & 3.19 (0.06) & 3.19 (0.06) & \textbf{3.15 (0.09)} & 3.19 (0.06) & 3.22 (0.15) \\
  WPI & 10 & \textbf{3.01 (0.05)} & \textbf{3.01 (0.05)} & 3.02 (0.05) & 3.06 (0.06) & \textbf{3.01 (0.05)} & 3.11 (0.11) \\
  WPI & 20 & \textbf{2.80 (0.07)} & \textbf{2.80 (0.07)} & \textbf{2.80 (0.07)} & 2.91 (0.08) & \textbf{2.80 (0.07)} & 2.94 (0.07) \\
\bottomrule
\end{tabular}
\end{table}

\begin{table}[t]
\centering
\caption{\textbf{Absolute score error} $|\hat\theta_t - \hat\theta|$, mean (SD), under $\mathcal{M}$-open simulation (100 replicates per ability).
Lower is better; \textbf{bold} = best mean per row.
F = Fisher; BF = Bayesian Fisher; GI = global information; BV = Bayesian variance; E = entropy (greedy); SE = entropy (stochastic).}
\label{tab:mopen_l2}
\small
\begin{tabular}{@{}ll r r r r r r@{}}
\toprule
Dataset & $\theta$ & F & BF & GI & BV & E & SE \\
\midrule
\multicolumn{8}{@{}l}{\textit{RWA (22 items, $K{=}9$), $t = 5$}} \\
  & low & 2.41 (0.19) & 2.41 (0.19) & 2.41 (0.19) & 2.48 (0.26) & 2.41 (0.19) & \textbf{2.19 (0.16)} \\
  & mid & 1.80 (0.10) & 1.80 (0.10) & 1.80 (0.10) & 1.76 (0.11) & 1.80 (0.10) & \textbf{1.69 (0.05)} \\
  & high & \textbf{1.71 (0.10)} & \textbf{1.71 (0.10)} & \textbf{1.71 (0.10)} & 1.76 (0.05) & \textbf{1.71 (0.10)} & 1.82 (0.06) \\
  & all & 1.96 (0.33) & 1.96 (0.33) & 1.96 (0.33) & 1.98 (0.37) & 1.96 (0.33) & \textbf{1.89 (0.23)} \\
\midrule
\multicolumn{8}{@{}l}{\textit{$t = 10$}} \\
  & low & \textbf{1.74 (0.12)} & \textbf{1.74 (0.12)} & \textbf{1.74 (0.12)} & 1.95 (0.18) & \textbf{1.74 (0.12)} & 1.77 (0.10) \\
  & mid & \textbf{1.36 (0.06)} & \textbf{1.36 (0.06)} & \textbf{1.36 (0.06)} & 1.37 (0.10) & \textbf{1.36 (0.06)} & 1.37 (0.06) \\
  & high & \textbf{1.43 (0.07)} & \textbf{1.43 (0.07)} & \textbf{1.43 (0.07)} & 1.45 (0.05) & \textbf{1.43 (0.07)} & 1.53 (0.11) \\
  & all & \textbf{1.50 (0.19)} & \textbf{1.50 (0.19)} & \textbf{1.50 (0.19)} & 1.57 (0.28) & \textbf{1.50 (0.19)} & 1.54 (0.19) \\
\midrule
\multicolumn{8}{@{}l}{\textit{$t = 20$}} \\
  & low & \textbf{1.08 (0.04)} & \textbf{1.08 (0.04)} & \textbf{1.08 (0.04)} & \textbf{1.08 (0.09)} & \textbf{1.08 (0.04)} & 1.15 (0.04) \\
  & mid & \textbf{0.83 (0.06)} & \textbf{0.83 (0.06)} & \textbf{0.83 (0.06)} & 0.85 (0.07) & \textbf{0.83 (0.06)} & 0.87 (0.07) \\
  & high & \textbf{0.90 (0.07)} & \textbf{0.90 (0.07)} & \textbf{0.90 (0.07)} & \textbf{0.90 (0.08)} & \textbf{0.90 (0.07)} & \textbf{0.90 (0.08)} \\
  & all & \textbf{0.93 (0.12)} & \textbf{0.93 (0.12)} & \textbf{0.93 (0.12)} & 0.94 (0.13) & \textbf{0.93 (0.12)} & 0.96 (0.14) \\
\midrule
\multicolumn{8}{@{}l}{\textit{NPI (40 items, $K{=}2$), $t = 5$}} \\
  & low & 1.29 (0.16) & 1.29 (0.16) & 1.29 (0.16) & \textbf{1.11 (0.10)} & 1.29 (0.16) & 1.15 (0.14) \\
  & mid & 1.07 (0.07) & 1.07 (0.07) & 1.07 (0.07) & \textbf{0.95 (0.08)} & 1.07 (0.07) & 0.96 (0.09) \\
  & high & \textbf{0.99 (0.11)} & \textbf{0.99 (0.11)} & \textbf{0.99 (0.11)} & 1.20 (0.18) & \textbf{0.99 (0.11)} & 1.22 (0.13) \\
  & all & 1.11 (0.17) & 1.11 (0.17) & 1.11 (0.17) & \textbf{1.07 (0.16)} & 1.11 (0.17) & 1.10 (0.16) \\
\midrule
\multicolumn{8}{@{}l}{\textit{$t = 10$}} \\
  & low & 1.12 (0.09) & 1.15 (0.10) & 1.14 (0.11) & \textbf{1.01 (0.09)} & 1.15 (0.10) & 1.05 (0.16) \\
  & mid & 0.92 (0.06) & 0.93 (0.06) & 0.93 (0.06) & \textbf{0.87 (0.08)} & 0.93 (0.06) & \textbf{0.87 (0.10)} \\
  & high & 0.84 (0.08) & 0.84 (0.05) & \textbf{0.83 (0.06)} & 1.05 (0.10) & 0.85 (0.06) & 1.04 (0.09) \\
  & all & \textbf{0.96 (0.14)} & 0.97 (0.15) & 0.97 (0.15) & 0.97 (0.12) & 0.97 (0.14) & 0.98 (0.15) \\
\midrule
\multicolumn{8}{@{}l}{\textit{$t = 20$}} \\
  & low & 0.77 (0.03) & 0.78 (0.03) & 0.78 (0.04) & 0.83 (0.08) & 0.78 (0.03) & \textbf{0.75 (0.10)} \\
  & mid & 0.69 (0.03) & 0.69 (0.03) & 0.68 (0.04) & 0.66 (0.05) & 0.69 (0.03) & \textbf{0.65 (0.04)} \\
  & high & \textbf{0.65 (0.07)} & \textbf{0.65 (0.07)} & \textbf{0.65 (0.07)} & 0.82 (0.13) & \textbf{0.65 (0.07)} & 0.74 (0.04) \\
  & all & \textbf{0.70 (0.07)} & 0.71 (0.07) & \textbf{0.70 (0.07)} & 0.76 (0.12) & 0.71 (0.07) & 0.71 (0.08) \\
\midrule
\multicolumn{8}{@{}l}{\textit{TMA (50 items, $K{=}2$), $t = 5$}} \\
  & low & 1.32 (0.09) & 1.31 (0.08) & 1.32 (0.09) & \textbf{1.20 (0.15)} & 1.31 (0.08) & 1.30 (0.14) \\
  & mid & 1.06 (0.04) & 1.06 (0.05) & 1.04 (0.03) & \textbf{0.91 (0.08)} & 1.06 (0.03) & 0.93 (0.08) \\
  & high & 1.02 (0.04) & \textbf{1.00 (0.05)} & \textbf{1.00 (0.04)} & 1.12 (0.18) & \textbf{1.00 (0.04)} & \textbf{1.00 (0.11)} \\
  & all & 1.13 (0.14) & 1.12 (0.15) & 1.12 (0.15) & \textbf{1.07 (0.19)} & 1.12 (0.14) & \textbf{1.07 (0.19)} \\
\midrule
\multicolumn{8}{@{}l}{\textit{$t = 10$}} \\
  & low & 1.05 (0.08) & 1.04 (0.08) & 1.04 (0.08) & \textbf{1.03 (0.10)} & 1.04 (0.08) & 1.11 (0.10) \\
  & mid & 0.91 (0.04) & 0.92 (0.05) & 0.91 (0.04) & 0.89 (0.06) & 0.91 (0.05) & \textbf{0.84 (0.05)} \\
  & high & \textbf{0.91 (0.04)} & \textbf{0.91 (0.06)} & 0.92 (0.05) & 1.06 (0.13) & \textbf{0.91 (0.05)} & \textbf{0.91 (0.03)} \\
  & all & \textbf{0.95 (0.09)} & \textbf{0.95 (0.08)} & \textbf{0.95 (0.08)} & 0.98 (0.13) & \textbf{0.95 (0.09)} & \textbf{0.95 (0.13)} \\
\midrule
\multicolumn{8}{@{}l}{\textit{$t = 20$}} \\
  & low & \textbf{0.69 (0.02)} & 0.70 (0.03) & 0.70 (0.03) & 0.86 (0.12) & 0.70 (0.03) & 0.83 (0.07) \\
  & mid & 0.69 (0.04) & 0.69 (0.03) & \textbf{0.68 (0.04)} & 0.69 (0.05) & 0.69 (0.04) & 0.72 (0.02) \\
  & high & \textbf{0.70 (0.03)} & \textbf{0.70 (0.03)} & \textbf{0.70 (0.03)} & 0.77 (0.07) & \textbf{0.70 (0.03)} & 0.73 (0.06) \\
  & all & \textbf{0.69 (0.03)} & \textbf{0.69 (0.03)} & \textbf{0.69 (0.04)} & 0.77 (0.11) & 0.70 (0.03) & 0.76 (0.07) \\
\midrule
\multicolumn{8}{@{}l}{\textit{EQSQ (120 items, $K{=}4$), $t = 5$}} \\
  & low & \textbf{1.01 (0.05)} & 1.07 (0.07) & 1.08 (0.08) & 1.27 (0.18) & 1.07 (0.07) & 1.14 (0.07) \\
  & mid & 0.90 (0.06) & 0.92 (0.08) & 0.92 (0.07) & \textbf{0.81 (0.06)} & 0.92 (0.08) & 0.84 (0.10) \\
  & high & 0.96 (0.08) & 0.98 (0.06) & 0.99 (0.06) & \textbf{0.80 (0.06)} & 0.97 (0.06) & 0.87 (0.07) \\
  & all & 0.95 (0.08) & 0.99 (0.09) & 0.99 (0.10) & 0.95 (0.24) & 0.98 (0.09) & \textbf{0.94 (0.16)} \\
\midrule
\multicolumn{8}{@{}l}{\textit{$t = 10$}} \\
  & low & \textbf{0.91 (0.06)} & 0.92 (0.05) & 0.92 (0.05) & 1.04 (0.14) & 0.92 (0.05) & 0.97 (0.05) \\
  & mid & 0.81 (0.04) & 0.82 (0.04) & 0.82 (0.03) & \textbf{0.79 (0.08)} & 0.82 (0.04) & 0.83 (0.05) \\
  & high & 0.84 (0.03) & 0.84 (0.04) & 0.84 (0.03) & \textbf{0.77 (0.06)} & 0.84 (0.04) & 0.82 (0.05) \\
  & all & \textbf{0.85 (0.06)} & 0.86 (0.06) & 0.86 (0.06) & 0.86 (0.15) & 0.86 (0.06) & 0.87 (0.08) \\
\midrule
\multicolumn{8}{@{}l}{\textit{$t = 20$}} \\
  & low & \textbf{0.70 (0.05)} & 0.71 (0.04) & 0.71 (0.04) & 0.85 (0.09) & 0.71 (0.05) & 0.78 (0.03) \\
  & mid & \textbf{0.66 (0.05)} & 0.67 (0.05) & 0.68 (0.05) & 0.67 (0.06) & 0.67 (0.05) & 0.68 (0.07) \\
  & high & 0.71 (0.05) & 0.71 (0.05) & 0.71 (0.05) & \textbf{0.63 (0.03)} & 0.71 (0.06) & 0.74 (0.04) \\
  & all & \textbf{0.69 (0.06)} & 0.70 (0.05) & 0.70 (0.05) & 0.71 (0.11) & 0.70 (0.05) & 0.73 (0.07) \\
\midrule
\multicolumn{8}{@{}l}{\textit{GRIT (12 items, $K{=}5$), $t = 5$}} \\
  & low & \textbf{1.58 (0.15)} & \textbf{1.58 (0.15)} & \textbf{1.58 (0.15)} & 2.12 (0.27) & \textbf{1.58 (0.15)} & 1.83 (0.17) \\
  & mid & \textbf{1.21 (0.05)} & \textbf{1.21 (0.05)} & \textbf{1.21 (0.05)} & 1.37 (0.14) & \textbf{1.21 (0.05)} & 1.33 (0.10) \\
  & high & 1.20 (0.10) & 1.20 (0.10) & 1.20 (0.10) & \textbf{1.18 (0.10)} & 1.20 (0.10) & 1.22 (0.13) \\
  & all & \textbf{1.32 (0.20)} & \textbf{1.32 (0.20)} & \textbf{1.32 (0.20)} & 1.54 (0.44) & \textbf{1.32 (0.20)} & 1.45 (0.29) \\
\midrule
\multicolumn{8}{@{}l}{\textit{$t = 10$}} \\
  & low & \textbf{1.26 (0.14)} & \textbf{1.26 (0.14)} & \textbf{1.26 (0.14)} & 1.41 (0.16) & \textbf{1.26 (0.14)} & 1.33 (0.10) \\
  & mid & \textbf{0.90 (0.08)} & \textbf{0.90 (0.08)} & \textbf{0.90 (0.08)} & 0.92 (0.07) & \textbf{0.90 (0.08)} & 0.93 (0.09) \\
  & high & 0.85 (0.06) & 0.85 (0.06) & 0.85 (0.06) & 0.82 (0.07) & 0.85 (0.06) & \textbf{0.81 (0.08)} \\
  & all & \textbf{1.00 (0.20)} & \textbf{1.00 (0.20)} & \textbf{1.00 (0.20)} & 1.04 (0.27) & \textbf{1.00 (0.20)} & 1.02 (0.23) \\
\midrule
\multicolumn{8}{@{}l}{\textit{SCS (10 items, $K{=}4$), $t = 5$}} \\
  & low & 1.63 (0.14) & 1.63 (0.14) & 1.63 (0.14) & \textbf{1.48 (0.18)} & 1.63 (0.14) & 1.63 (0.15) \\
  & mid & 1.30 (0.07) & 1.30 (0.07) & 1.30 (0.07) & \textbf{1.22 (0.09)} & 1.30 (0.07) & 1.29 (0.08) \\
  & high & 1.38 (0.09) & 1.37 (0.09) & 1.37 (0.09) & 1.41 (0.13) & 1.38 (0.09) & \textbf{1.36 (0.10)} \\
  & all & 1.42 (0.17) & 1.42 (0.17) & 1.42 (0.17) & \textbf{1.36 (0.18)} & 1.42 (0.17) & 1.42 (0.18) \\
\midrule
\multicolumn{8}{@{}l}{\textit{$t = 10$}} \\
  & low & \textbf{1.06 (0.10)} & \textbf{1.06 (0.10)} & \textbf{1.06 (0.10)} & \textbf{1.06 (0.10)} & \textbf{1.06 (0.10)} & \textbf{1.06 (0.10)} \\
  & mid & \textbf{0.84 (0.05)} & \textbf{0.84 (0.05)} & \textbf{0.84 (0.05)} & \textbf{0.84 (0.05)} & \textbf{0.84 (0.05)} & \textbf{0.84 (0.05)} \\
  & high & \textbf{0.91 (0.07)} & \textbf{0.91 (0.07)} & \textbf{0.91 (0.07)} & \textbf{0.91 (0.07)} & \textbf{0.91 (0.07)} & \textbf{0.91 (0.07)} \\
  & all & \textbf{0.93 (0.12)} & \textbf{0.93 (0.12)} & \textbf{0.93 (0.12)} & \textbf{0.93 (0.12)} & \textbf{0.93 (0.12)} & \textbf{0.93 (0.12)} \\
\midrule
\multicolumn{8}{@{}l}{\textit{GCBS (15 items, $K{=}5$), $t = 5$}} \\
  & low & \textbf{1.61 (0.09)} & \textbf{1.61 (0.09)} & \textbf{1.61 (0.08)} & 1.80 (0.17) & \textbf{1.61 (0.08)} & 1.71 (0.16) \\
  & mid & \textbf{1.35 (0.08)} & \textbf{1.35 (0.09)} & \textbf{1.35 (0.09)} & 1.48 (0.08) & \textbf{1.35 (0.09)} & 1.48 (0.05) \\
  & high & 1.68 (0.14) & 1.69 (0.14) & 1.69 (0.14) & 1.64 (0.21) & 1.69 (0.14) & \textbf{1.63 (0.20)} \\
  & all & \textbf{1.53 (0.18)} & \textbf{1.53 (0.18)} & \textbf{1.53 (0.18)} & 1.63 (0.20) & \textbf{1.53 (0.18)} & 1.60 (0.18) \\
\midrule
\multicolumn{8}{@{}l}{\textit{$t = 10$}} \\
  & low & 1.23 (0.09) & \textbf{1.22 (0.09)} & \textbf{1.22 (0.09)} & 1.39 (0.13) & 1.23 (0.09) & 1.28 (0.12) \\
  & mid & \textbf{1.07 (0.05)} & \textbf{1.07 (0.04)} & \textbf{1.07 (0.04)} & 1.16 (0.05) & \textbf{1.07 (0.04)} & 1.10 (0.05) \\
  & high & 1.23 (0.11) & 1.22 (0.11) & 1.22 (0.12) & \textbf{1.20 (0.17)} & 1.23 (0.11) & 1.22 (0.19) \\
  & all & 1.17 (0.11) & \textbf{1.16 (0.11)} & \textbf{1.16 (0.11)} & 1.24 (0.16) & \textbf{1.16 (0.11)} & 1.19 (0.15) \\
\midrule
\multicolumn{8}{@{}l}{\textit{WPI (116 items, $K{=}2$), $t = 5$}} \\
  & low & 1.08 (0.04) & 1.10 (0.04) & 1.10 (0.03) & \textbf{0.81 (0.08)} & 1.10 (0.04) & 0.84 (0.03) \\
  & mid & 1.06 (0.06) & 1.05 (0.08) & 1.07 (0.07) & \textbf{0.81 (0.05)} & 1.05 (0.08) & 1.01 (0.13) \\
  & high & 1.13 (0.07) & \textbf{1.11 (0.07)} & 1.12 (0.07) & 1.26 (0.18) & \textbf{1.11 (0.07)} & 1.48 (0.19) \\
  & all & 1.09 (0.07) & 1.08 (0.07) & 1.09 (0.07) & \textbf{0.95 (0.23)} & 1.08 (0.07) & 1.10 (0.30) \\
\midrule
\multicolumn{8}{@{}l}{\textit{$t = 10$}} \\
  & low & 0.94 (0.06) & 0.95 (0.05) & 0.95 (0.05) & 0.85 (0.06) & 0.94 (0.05) & \textbf{0.83 (0.03)} \\
  & mid & 0.87 (0.03) & 0.86 (0.04) & 0.88 (0.04) & \textbf{0.81 (0.01)} & 0.87 (0.04) & 0.95 (0.06) \\
  & high & 0.92 (0.05) & \textbf{0.91 (0.06)} & 0.92 (0.05) & 1.01 (0.17) & 0.92 (0.05) & 1.20 (0.12) \\
  & all & 0.91 (0.06) & 0.90 (0.06) & 0.92 (0.05) & \textbf{0.89 (0.13)} & 0.91 (0.06) & 0.99 (0.17) \\
\midrule
\multicolumn{8}{@{}l}{\textit{$t = 20$}} \\
  & low & 0.78 (0.06) & 0.77 (0.04) & 0.76 (0.05) & \textbf{0.71 (0.03)} & 0.77 (0.05) & 0.79 (0.04) \\
  & mid & 0.74 (0.07) & 0.74 (0.07) & 0.74 (0.06) & \textbf{0.70 (0.04)} & 0.74 (0.07) & 0.84 (0.07) \\
  & high & \textbf{0.73 (0.04)} & \textbf{0.73 (0.04)} & \textbf{0.73 (0.04)} & 0.93 (0.13) & \textbf{0.73 (0.04)} & 0.93 (0.05) \\
  & all & \textbf{0.75 (0.06)} & \textbf{0.75 (0.05)} & \textbf{0.75 (0.06)} & 0.78 (0.13) & \textbf{0.75 (0.06)} & 0.85 (0.08) \\
\bottomrule
\end{tabular}
\end{table}

\end{document}